\documentclass[fleqn,twoside,twocolumn]{revtex4} 
\usepackage[sec]{ujp} 

\begin{document}
\title[OSCILLATORY RULE IN THE ENERGY SPECTRUM]
{OSCILLATORY RULE IN THE ENERGY SPECTRUM\\ OF TRAPS IN KCl AND NaI CRYSTALS}%
\author{A.F. Gumenyuk}
\author{S.Yu. Kutovyi}%
\author{O.P. Stanovyi}%
\author{V.G.~Pashchenko,~S.V.~Tomylko}%
\affiliation{\knu}%
\address{2, Academician Glushkov Prosp., Kyiv 03127}%
\email{lns@univ.kiev.ua}
\avtorcol{A.F. GUMENYUK, S.Yu. KUTOVYI, O.P. STANOVYI et al.}
\udk{535.377; 535.375} \pacs{78.60.XXXX} \razd{\secvii}

\setcounter{page}{999}%
\maketitle

\begin{abstract}
The thermoluminescence (TL) method is used for the investigation
of the energy spectrum of traps in KCl and NaI crystals in the
temperature range 80--500 K. It is shown that the thermal activation energies of traps in KCl and NaI
form one oscillatory series $E_{n}=\hbar\omega n$ with vibrational
quantum energies of 0.121 eV (979 cm$^{-1}$) in KCl and 0.061 eV
(492 $cm^{-1}$) in NaI. In this case, the quantum number \emph{n}
assumes half-integer and integer values. Based on the generalized
data on the investigated alkali-halide crystals (AHC), we
confirmed the earlier proposed model of TL in AHCs. It is assumed
that, in addition to the nonradiative \emph{H-F} recombination, there
exists the two-stage recombination of \emph{H}-centers at anion
vacancies resulting in the radiative recombination of a hole at an
\emph{F}-center. The energy of a quantum in the oscillatory rule
corresponds to a local vibrational mode of an $X_{2}^{-}$ halide
molecule.
\end{abstract}

\section{Introduction}

Investigating the energy spectrum of traps in a number of crystals
(about 15) with ion-covalent bonds, we discovered [1--11] that the
thermal activation energy of traps is described by the harmonic
oscillator formula $E_{n}= \hbar\omega_{\mathrm{TL}}(n+1/2)$, where
the quantum energy $\hbar\omega_{\mathrm{TL}}$ assumes values
characteristic of lattice vibrations (0.01--0.18 eV, i.e.
$\sim100-1500$ cm$^{-1}$). We observed from one to five oscillatory
series, each of which is characterized by its vibrational quantum
energy $\hbar\omega_{\mathrm{TL}}$. Except for alkali halide
crystals (AHC), the majority of the values of
$\hbar\omega_{\mathrm{TL}}$ coincide with the energies of
high-frequency lines of the basic first-order Raman scattering (RS)
spectrum $\hbar\omega_{\mathrm{RS}}$, whereas the other values of
$\hbar\omega_{\mathrm{TL}}$ correspond to the energies of lines not
predicted by the group-theoretical analysis (second-order RS lines,
local vibrational modes, lines forbidden by RS selection rules). No
correspondence is found between $\hbar\omega_{\mathrm{TL}}$ and
$\hbar\omega_{\mathrm{RS}}$ for crystals with simple lattices --
NaCl and LiF AHCs. The obtained results gave grounds to state that
the oscillatory rule reflects some general regularity rather than
represents a unique peculiarity of certain crystal compounds. A
polaron model of traps and TL process was proposed
in~[10].\looseness=1

In cubic AHCs, there are no vibrational modes  active in RS, though
the oscillatory rule for the trap energies with the vibrational quantum
energies $\hbar\omega_{\mathrm{NaCl}}=0.112$ eV and
$\hbar\omega_{\mathrm{LiF}}=0.162$ eV is also observed in the investigated NaCl and LiF crystals. These values
exceed the energy of the only optical vibrational mode in these
crystals by several times. We did not find any manifestations of the indicated
frequencies in other phenomena, though we indicate the correlation between
these frequencies and the masses of halogen atoms:
$\hbar\omega_{\mathrm{NaCl}}/\hbar\omega_{\mathrm{LiF}}=1.44$ and
$\sqrt{m_{\mathrm{Cl}}/m_{F}}=1.37$. This fact stimulated our
searches for the oscillatory rule and the corresponding correlation
in other AHCs. For the further studies, we chose KCl crystals, where
we expected to find a vibrational frequency close to
$\hbar\omega_{\mathrm{NaCl}}$, as well as NaI, whose vibrational
frequency was predicted to be lower than that in NaCl by a factor of
$\sim2$. The work describes the results of these experiments and the
generalizing conclusions concerning the TL mechanisms in
AHCs.

\section{Experimental Technique}

The investigations were performed in the temperature range
80--500~K. The heating rate was $0.2\pm5\%$~K/s. The linear increase
of the temperature and the data registration were controlled by a
computer program. Samples 0.2--0.5 mm in thickness and 4--8~mm in
transverse size were located in a vacuum cryostat and excited by
X-radiation (BSV-2, W, 30~kV, 10 mA) through a beryllium window 0.5
mm in thickness.

The samples were cut off from KCl crystals both  specially undoped
and Eu-doped. In the doped crystals, the TL peaks appeared to be
rather intense and split, which allowed us to investigate the energy
spectrum of traps using the fractional technique described in our
previous works. KCl:Mg,Ca crystals appeared to be inappropriate for
studying TL with the help of the fractional technique, because the
corresponding peaks were substantially broadened due to the
considerable concentration of impurities $10^{17}-10^{18}$
cm$^{-3}$, and the trap energies were no more discrete. Some results
were obtained for undoped KCl crystals both synthetic and natural.
They are less accurate as compared to those obtained for KCl:Eu
crystals, as our samples of undoped KCl manifested much weaker TL
than doped crystals.

As another material, we used dosimetric NaI:Tl crystals both
without  preliminary thermal treatments and those annealed under
various conditions (variation of the anneal temperature and
duration with the following fast cooling). The change of the
anneal parameters was used to increase the intensity of individual
maxima of the heat-treated samples, which improved the sensitivity
of fractional experiments.

The experimental procedures and the technique of data treatment
were considered in our previous works [5--8] in detail.

\section{Experimental Results}

\subsection*{3.1. KCl}

\subsubsection*{General description of TL curves}

Figure 1 shows the TL curves  of some investigated samples without
preliminary thermal treatment both undoped ({\it a}) and Eu-doped
({\it b}).

\begin{figure*}
\includegraphics[width=8.5cm]{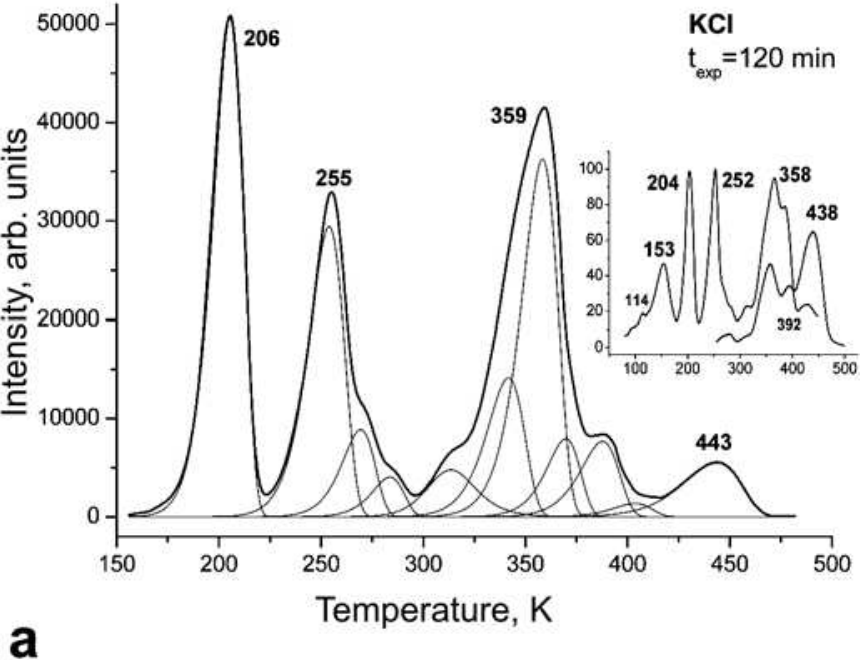}\hspace{0.5cm}\includegraphics[width=8.3cm]{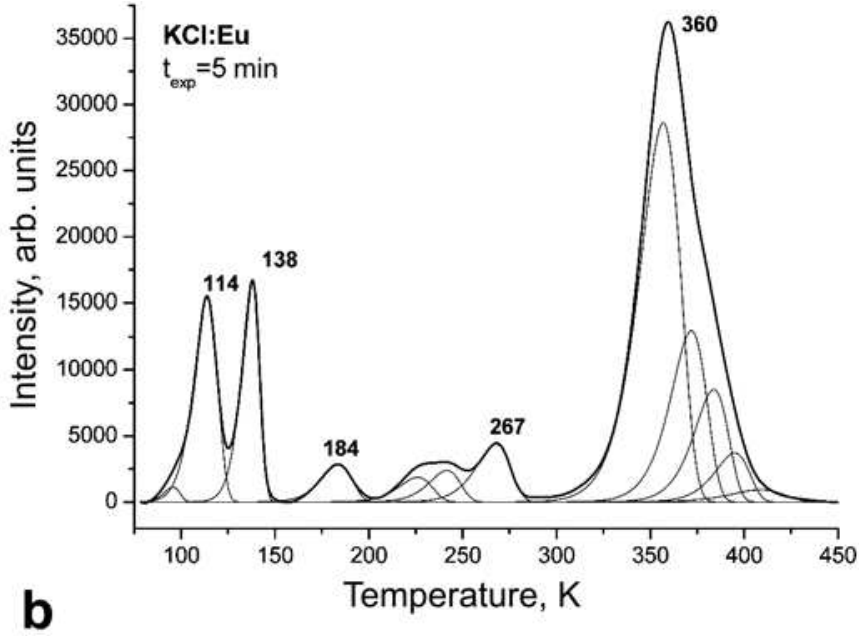}
\vskip-3mm\caption{TL curves of KCl ({\it a}) and KCl:Eu ({\it b})
crystals and their decomposition into elementary contours }
\end{figure*}

The TL curve of the undoped sample includes four intense bands with
maxima  at 206, 255, 359, and 443 K. The TL curves were decomposed
into elementary contours. For the sake of agreement with the
experimental curves, each contour was chosen in the form
corresponding to the linear and quadratic kinetics [12]. It turns
out (Fig. 1,$a$) that the 206-K peak is elementary, whereas the
255-K one consists of three components. Moreover, the intensity of
the principal component (254 K) considerably exceeds that of two
high-temperature satellites at 269 and 284 K. The 359-K peak
includes an intense peak at 358 K and four relatively weak ones at
314, 342, 370, and 387 K. The peak at 443 K is elementary. All the
indicated peaks glow according to the linear kinetics. In addition,
the temperatures of their maxima practically do not depend on the
irradiation dose, which also testifies to the linear kinetics of the
process.

A more pronounced overlapping observed for the peaks at higher
temperatures (caused by the proportional increase of their width)
considerably complicates the procedure of fractional glowing of
individual peaks. Therefore, the accuracy of determination of their
activation energies significantly worsens, as the program used for
the calculation of activation energies based on initial slopes of TL
curves describes the glowing of a fraction with the help of only one
exponent that must fit into a straight line in the graph
$ln(I/n)=f(1/T)$ with a minimal deviation. This fact explains the
increasing spread of the trap activation energies in the case of
more high-temperature peaks -- if a peak is compound, then the
calculation results in a relatively larger spread of the activation
energies at a shelf (as compared to a simple peak) even in the case
of the domination of one of the components. However, due to the fact
that the lateral peaks in KCl are of considerably lower intensity,
we still used such a way of calculation of the energies.

The TL curves of the other samples of undoped KCl include a band of
low-temperature TL peaks in addition to the described ones (the
general maximum at 153 K), while the peaks at 314, 387, and 443 K
are more pronounced and have a higher intensity (inset in Fig.
1,{\it a}).

The decomposition of the TL curve of doped KCl:Eu into elementary
contours (Fig. 1,$b$) demonstrates that it consists of a compound
peak at 114~K, two simple ones at 138 and 184~K, as well as bands at
220--270~K and 320--420~K. Moreover, the peaks at 114, 138, and 184
K can be considered elementary (the low-temperature \mbox{95-K}
component is negligibly weak), the band at 220--270~K consists of
three simple peaks (226, 241, and 268~K) of approximately equal
intensities, while the intense band at 320--420~K consists of four
components (357, 372, 384, and 395 K), where the most intense peak
is the 357-K one (the same way as in undoped KCl). The glowing
kinetics of these peaks is also linear.

In order to obtain information on the nature (intrinsic or impurity)
of traps, we investigated thermoluminescence of undoped KCl and
KCl:Eu depending on the X-ray irradiation dose in the range 1--180
min at the nitrogen temperature. It turns out that the light sum
stored in each peak linearly increases with excitation time in the
whole investigated time interval. The only exclusion is the peak at
$\sim268$ K, whose intensity saturates after the 120-min irradiation
both in doped and in undoped KCl, so this peak is not related to Eu
impurity.\looseness=1

The linear dependence on the irradiation dose, i.e. the constant
rate of storage of the light sum testifies to the fact that the trap
population practically is not changed with increasing the time of
X-ray irradiation. Such a dependence can be explained taking into
account the local nature of X-ray excitation. A high-energy electron
generates free electrons and holes as well as interstitial ions and
vacancies with a considerable concentration of $10^{18}-10^{20}$
cm$^{-3}$ in a small region of the crystal. An optical excitation
source of the same electric power has a much larger photon flux
density (by 3--4 orders of magnitude). Each absorbed optical photon
can release only one charge, whereas a photoelectron liberated by an
X-ray quantum generates a substantial number of excitations located
in a small volume. In addition, due to the inessential absorption
coefficient of X-radiation in a substance consisting of light
elements ($\sim10$ cm$^{-1}$), excitation regions occupy the whole
sample, while optical radiation from the fundamental region ($>100$
cm$^{-1}$) excites only a thin ($\sim0.01$ cm) surface layer. As a
result, every following X-ray quantum is mainly absorbed in the
unexcited region of the sample, and one will observe a proportional
dependence of the light sum on the excitation time until excitation
regions start to overlap. It is worth noting that, in
Y$_{3}$Al$_{5}$O$_{12}$ crystals under similar excitation
conditions, one observes a sublinear dose dependence after the
20-minute excitation [13]. However, X-ray bremsstrahlung does not
generate radiation defects in this crystal. In this case, processes
of charge exchange of defects are determined by electrons and holes.
Due to their considerable mobility, they form much larger excitation
regions than those in alkali halides, as the dominant role in the
latter is played by radiationally created vacancies and interstices
with much smaller mobilities. Due to this fact, the overlapping of
excitation regions in KCl can be neglected even after the three-hour
irradiation.\looseness=-1

\subsubsection*{Fractional glowing. Oscillatory rule for trap\\
energies}

Experimental $E_{\exp}$ and calculated $E_{\mathrm{clc}}$
activation energies, maximum temperatures $T_{m}$, and frequency
factors $p_{0}$ are given in Table 1.

It is worth noting that the literature data on trap energies  in KCl
are very ambiguous (see, e.g., [14, 15]).

\begin{table}[b]
\noindent\caption{KCl: \boldmath$n$ is the quantum number;
\boldmath$E_{\exp}$ are the  experimental energies;
\boldmath$E_{\mathrm{clc}}$ are the energies calculated according to
the formula \boldmath$E_{n}=\hbar\omega n$,
\boldmath$\hbar\omega=0.121$ eV; \boldmath$T_{m}$ is the temperature
of the TL peak; \boldmath$p_{0}$ is the frequency
factor}\vskip3mm\tabcolsep10.6pt

\noindent{\footnotesize\begin{tabular}{c c c c c}
 \hline \multicolumn{1}{c}
{\rule{0pt}{9pt}$n$} & \multicolumn{1}{|c}{$E_{\mathrm{clc}}$}&
\multicolumn{1}{|c}{$E_{\exp}$}& \multicolumn{1}{|c}{$p_{0}$}&
\multicolumn{1}{|c}{$T_{m}$}\\%
\hline%
  1&0.121&$0.127\pm0.004$&$1.1\times 10^{4}$&114\\
  2&0.242&$0.248\pm0.005$&$2.0\times 10^{7}$&153\\
  3&0.363&$0.364\pm0.013$&$6.6\times 10^{11}$&138\\
  4&0.484&$0.484\pm0.006$&$3.9\times 10^{11}$&184\\
   &     &$0.475\pm0.006$&$2.7\times 10^{11}$&226\\
4.5&0.545&$0.546\pm0.002$&$5.2\times 10^{11}$&206\\
  5&0.605&$0.605\pm0.002$&$5.9\times 10^{10}$&241\\
  6&0.726&--              &--                &--\\
6.5&0.786&$0.770\pm0.007$&$1.7\times 10^{13}$&255\\
  7&0.847&$0.849\pm0.004$&$1.6\times 10^{14}$&268\\
$\ldots$&     &               &                 &   \\
9.5&1.149&$1.143\pm0.014$&$4.4\times 10^{14}$&360\\
 10&1.210&$1.211\pm0.007$&$2.5\times 10^{15}$&359\\
\hline
\end{tabular}}
\end{table}

\begin{figure*}
\includegraphics[width=8.2cm]{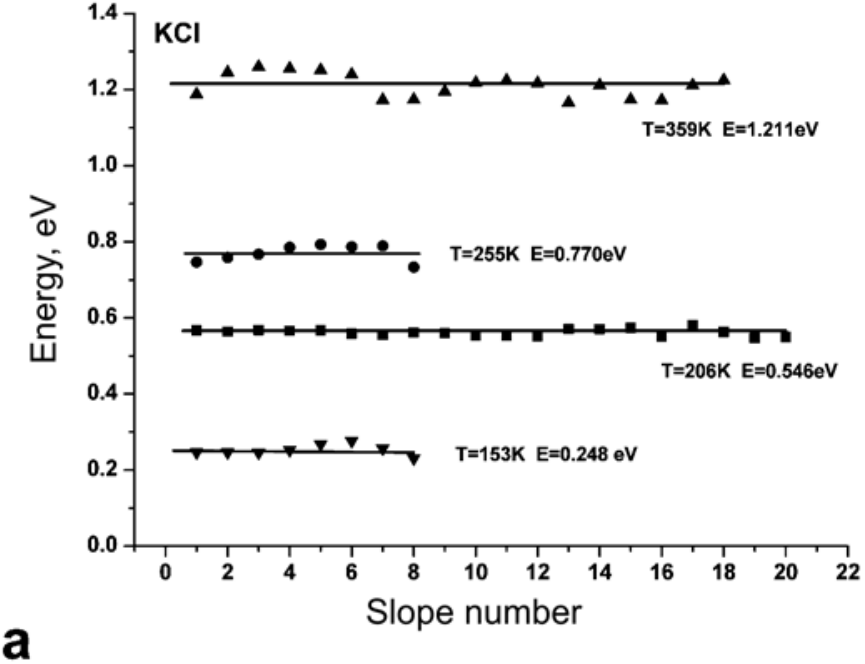}\hspace{0.5cm}\includegraphics[width=8.3cm]{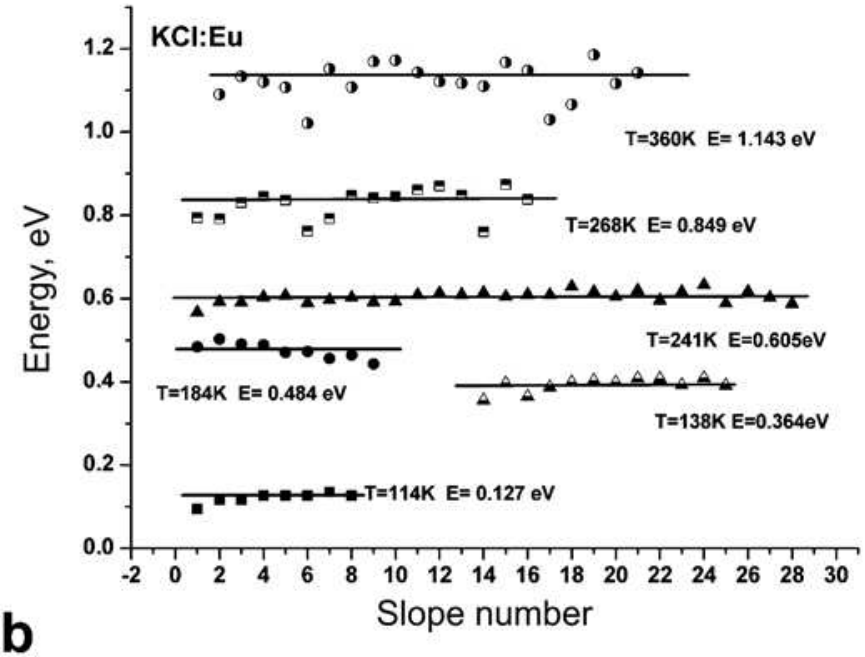}
\vskip-3.5mm \caption{Fractional deexcitation of KCl ({\it a}) and
KCl:Eu ({\it b}) }\vskip1mm
\end{figure*}

\begin{figure*}
\includegraphics[width=8.3cm]{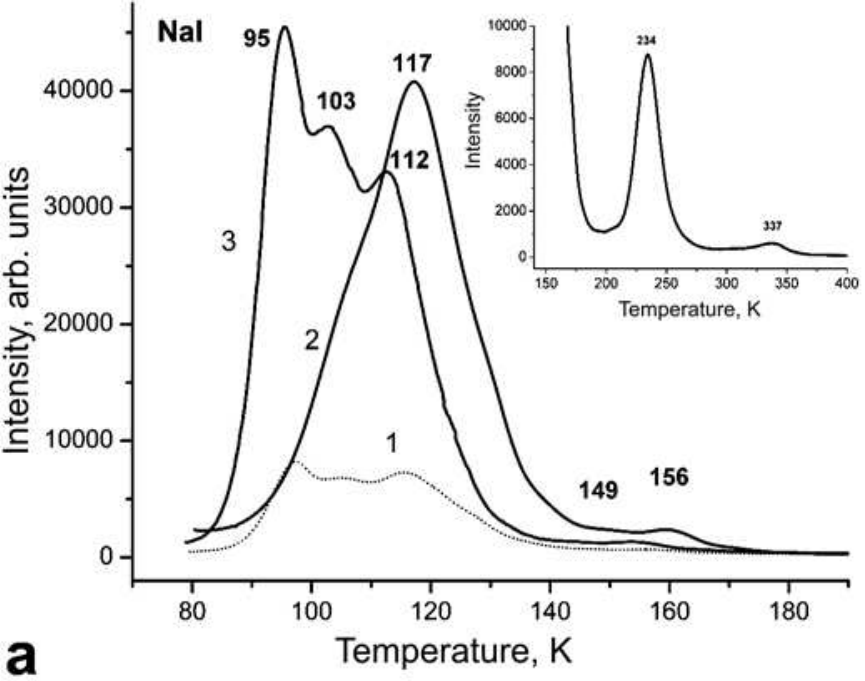}\hspace{0.5cm}\includegraphics[width=8.4cm]{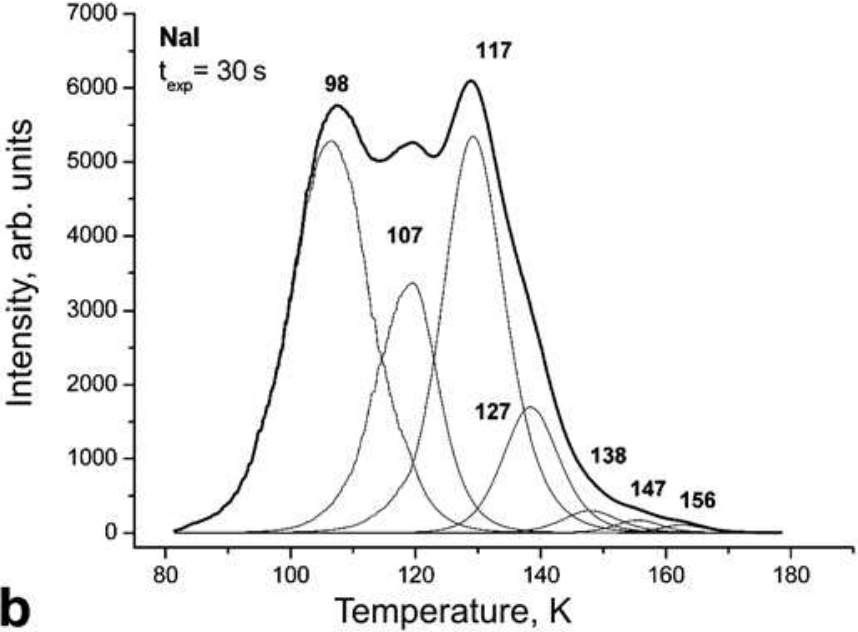}
\vskip-3.5mm\caption{TL curves of low-temperature NaI:Tl peaks at
different excitation times ({\it 1} -- 40 \emph{s}, {\it 2} -- 300
\emph{s}, hardened, {\it 3} -- 300 \emph{s}, annealed) and
high-temperature ones (inset)({\it a}); decomposition of the NaI:Tl
TL curve into elementary contours (30-second excitation) ({\it b})
}
\end{figure*}

In addition to Table 1, we adduce some results of calculating the
energies based on the fractional data in Fig.~2,{\it a} (undoped
KCl) and 2,$b$ (KCl:Eu), where the activation energies are presented
depending on the serial number of the fractional curves. The
activation energies $E_{\exp}$ were determined from the shelves of
these dependences. In order to make perception easier, the serial
numbers of the slopes for some shelves were increased by several
unities ($\sim10$). The error of determination of the energy was
estimated as the root-mean-square deviation of the fractional
energies at the shelf. The registration of all peaks is accompanied
by the background glow, whose intensity decreases with increasing
the number of the fractional curves and disappears for the last of
them; this background was eliminated with the help of the procedure
described in detail in [7--9].

As one can see from the table, all experimental energies are in good
agreement with the generalized oscillatory formula
$E_{n}=\hbar\omega n$, i.e. for integer and half-integer values of
$n$ (column $E_{\mathrm{clc}}$ of the table), the quantum energy
$\hbar\omega=0.121\pm0.004$ eV (979 cm$^{-1}$). The majority of trap
energies in KCl are multiple of $\hbar\omega$, except for the peaks
at 206 K (\emph{n}=4.5), 255 K (\emph{n}=6.5), and 360 K
(\emph{n}=9.5). The experiments on the fractional glowing for some
low-intensity peaks did not provide reliable results and therefore
were not included in Table 1.

\subsection*{3.2. NaI}

\subsubsection*{General description of TL curves} Figure 3,$a$
presents TL curves in the temperature range 80--180 K for NaI:Tl
crystal preirradiated by X-rays during 40 \emph{s} (curve {\it 1})
and 300 \emph{s} ({\it 2, 3}). The low-temperature band is compound;
one can distinguish at least seven close strongly overlapped peaks
of different intensities. The inset in Fig. 3,{\it a} shows the TL
curve for the region of higher temperatures. It contains two more
compound low-intensity TL peaks with the maximum temperatures of 234
and 337 K.

\begin{table}[b]
\noindent\caption{NaI:  \boldmath$n$ is the quantum number;
\boldmath$E_{\exp}$ are the experimental  energies;
\boldmath$E_{\mathrm{clc}}$ are the energies calculated according to
the formula \boldmath$E_{n}=\hbar\omega n$,
\boldmath$\hbar\omega=0.061$ eV; \boldmath$T_{m}$ is the maximum
temperature of the TL peak; \boldmath$p_{0}$ is the frequency
factor}\vskip3mm\tabcolsep7.2pt

\noindent{\footnotesize\begin{tabular}{c c c c c c}
 \hline \multicolumn{1}{c}
{\rule{0pt}{9pt}$n$} & \multicolumn{1}{|c}{$E_{\mathrm{clc} }$}&
\multicolumn{1}{|c}{$E_{\exp}$}& \multicolumn{2}{|c}{$p_{0}$ }&
\multicolumn{1}{|c}{$T_{m}$}\\\cline{4-5}%
\multicolumn{1}{c} {} & \multicolumn{1}{|c}{}&
\multicolumn{1}{|c}{}& \multicolumn{1}{|c}{eV}&
\multicolumn{1}{|c}{s$^{-1}$}&
\multicolumn{1}{|c}{$T_{m}$}\\%
\hline%
4.5&0.276&$0.284\pm0.004$&0.309 & $7.5\times 10^{13}$& 98\\
   5&0.305&$0.306\pm0.003$&0.273 & $6.6\times 10^{13}$&107\\
 5.5&0.336&$0.336\pm0.001$&0.279 & $6.7\times 10^{13}$&117
 \\
   6&0.366&$0.366\pm0.005$&0.598 & $1.4\times 10^{14}$&127\\
 $\ldots$&     &               & &                        &   \\
 7.5&0.458&$0.454\pm0.006$&0.432 & $1.0\times 10^{14}$&147\\
   8&0.488&$0.493\pm0.002$&1.697 & $4.1\times 10^{14}$&156\\
 $\ldots$&     &               &  &                       &   \\
  10&0.610&$0.615\pm0.013$&0.002 & $4.7\times 10^{11}$&234\\
 $\ldots$&     &               &  &                       &   \\
12.5&0.763&$0.764\pm0.009$&3.92  & $9.5\times 10^{14}$&234\\ \hline
\end{tabular}}
\end{table}

The form of the TL curve changes depending on the excitation time.
In general, the TL properties of NaI appeared to be very unstable
in the sense of the reproducibility of low-temperature TL curves
as compared to other studies of AHCs. Certain difficulties were
also induced by the hygroscopicity of the crystal. A positive
feature of this instability consists in the fact that each
following experiment was actually performed with a sample that
could be considered thermally treated. As it was shown in a number
of our earlier investigations, a considerable range of \emph{n} in
the course of searches for the oscillatory rule in the absence of
a set of samples grown under different conditions and doped by
different impurities was provided due to different conditions of
the thermal treatment. In addition, we used a dosimetric NaI:Tl
material, where intense thermoluminescence was observed.

The decomposition of the TL curve into elementary contours (Fig.
3,$b$)  demonstrates that the low-temperature band consists of seven
peaks: four ones are rather intense (98, 107, 117, and 127 K),
the other three peaks (138, 147, and 156 K) are weak. The form of all the
peaks corresponds to the quadratic kinetics.

The investigation of the dose dependences (as well as the decomposition
into contours)  shows that, for annealed samples, an increase of the
irradiation dose results, first of all, in the growth of the intensity
of the first low-temperature peak $\sim98$ K, whereas, for hardened
ones -- of the third 117-K peak (in this case, the intensity of the
98-K peak remains constant). Moreover, annealing or hardening appeared
to be equivalent to slow cooling and settling of a sample
($\sim10-12$~{hours}) or fast nitrogen cooling down to initial
temperatures. As a result, two successive records of TL curves
(obtained with a minimal pause) at the five-minute excitation gave very
different forms of the curves (Fig. 3,{\it a}, curves {\it 2, 3}),
whereas the forms of those recorded with an interval of twenty-four
hours were practically identical. With increasing the dose, the maxima
of the peaks for annealed samples shift toward lower temperatures
(down to 4~K), while such a shift for hardened ones is practically
absent ($<1$~K).

The quadratic kinetics is realized in the case of the recapture and
at approximately equal concentrations of carriers at traps
of a given kind and free luminescence centers (LC).
This is possible in the presence of genetically related low-mobility
pairs of traps and LCs.

\begin{figure}
\includegraphics[width=\column]{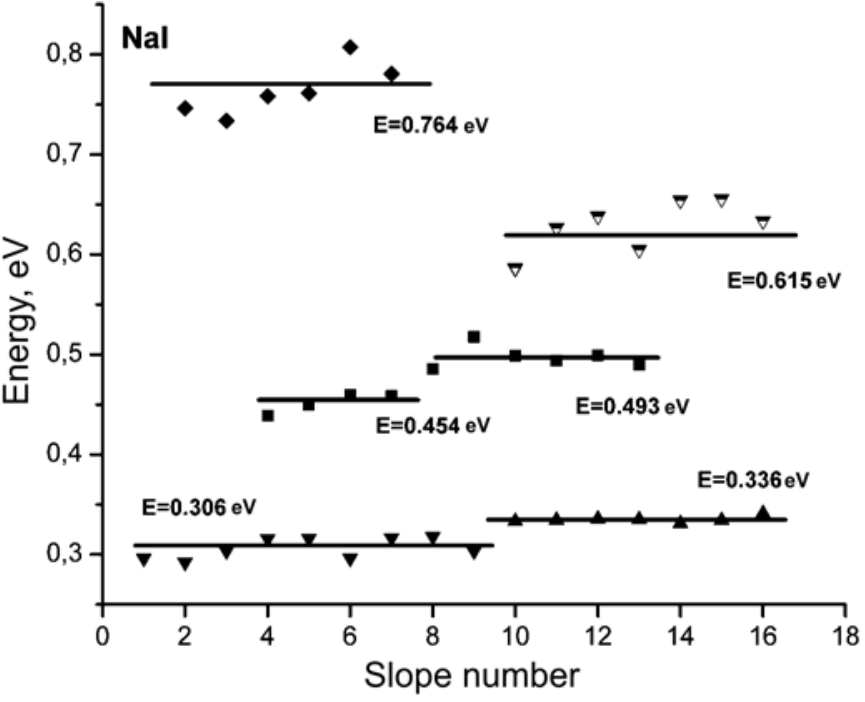}
\caption{Fractional glowing of NaI:Tl}
\end{figure}

\subsubsection*{Fractional glowing}

A considerable overlapping complicates the procedure of fractional
glowing and therefore substantially worsens the accuracy of
determination of the activation energy of each peak. In this case, a
partial clipping is ineffective, but we applied it as it still gave
some useful information. Some results of calculating the energies
based on fractional data are presented in Fig. 4 depending on the
serial number of the fractional curves. The values of $E_{\exp}$,
maximum temperatures $T_{m}$, and frequency factors $p_{0}$ are
given in Table 2. Due to the considerable overlapping of the
low-temperature TL peaks, it appeared impossible to definitely
correlate the activation energies with specific peaks from
fractional experiments. That is why such a correlation was performed
based on the correspondence of the frequency factor in the units of
eV to the value of one of the found energies. These values must be
close to each other. Our calculations demonstrated that such a
correlation could be performed rather unambiguously -- a variation
of the activation energy by $\sim0.02$ eV resulted in the change of
the frequency factor by orders of magnitude. A substantial
disagreement between the activation energy and the frequency factor
was observed only for the high-temperature peak at 234 K (nonannealed sample).

The same way as in the case of KCl,  the experimentally determined
energies agree with the oscillatory formula (column
$E_{\mathrm{clc}}$ of the table) for the quantum energy
$\hbar\omega_{\mathrm{TL}}=0.061$ eV (492 cm$^{-1}$). The totality
of the low-temperature peaks is characterized by a collection of
activation energies from 0.284 to 0.493 eV, and these energies fit
 the oscillatory rule with integer and half-integer \emph{n}
(\emph{n}=4.5--8). The activation energy of the 234-K peak
determined for the annealed sample is $0.615\pm0.013$ eV
(\emph{n}=10), for the nonannealed one -- $0.764\pm0.009$ eV
(\emph{n}=12.5). The results of the fractional glowing of the 337-K
peak appeared unreliable, that is why they are not included in the
table.

\section{Discussion of the Results}

\subsection*{4.1. Polaron model} The earlier obtained results
[1--11] allowed one to propose a model considering a filled trap as
a polaron that is self-trapped or stabilized by the local field of
an intrinsic defect or impurity. In all crystals not belonging to
AHCs, the vibrational TL frequencies correspond to modes of the
Raman scattering spectrum. In the case of AHCs including KCl and
Na${}^2$, there appear the local vibrational modes that form one-series
oscillatory rules. Unlike the materials we investigated earlier,
where the trap energy in a series was described by the formula
$E_{n}=\hbar\omega_{\mathrm{TL}}(n+1/2)$, \emph{n}=0,1,2..., the
energies of series in AHCs are generally described by the formula
$E_{n}=\hbar\omega_{\mathrm{TL}} n'$, where \emph{n'} can assume
both half-integer and integer values. For example, for traps glowing
in NaCl in the temperature range 80--500 K, one observes a
``classical'' oscillatory rule excluding one trap, whose energy
appeared to be multiple of the quantum number. In contrast, the
energies of all traps in LiF in this temperature region appear to be
multiple of the corresponding quantum number except for one trap,
whose energy is multiple of the half-integer quantum number. In KCl,
the energies of eight traps are multiple of the vibrational quantum
energy, whereas three of them correspond to its half-integer value
(Table 1). Na${}^{2}$ represents an intermediate variant: four trap
energies are described by the formula for half-integer \emph{n} and
the same number -- for integer one (Table 2).

The mechanism of existence of trap energies multiple of integer or
half-integer values of the quantum number \emph{n} was clarified in
our previous work [10]. Here, we briefly state the basic points. The
thermoactivation processes are accompanied by changes in the ion
subsystem that consist in a shift of a polaron in the field of some
defect -- a stabilization center. Due to the considerable local
polarization, this shift is realized by means of thermoactivation
jumps. A polaron trap can be presented in the energy diagram as a
parabolic potential well with equidistant vibrational levels. The
depth of the potential well, i.e. the number of the vibrational
level \emph{n}, from which a charge is thermally released, is
determined by the nature of the defect localizing the polaron. A
thermally released polaron can reach a luminescence center in the
following ways: a) according to the band mechanism realized in the
case where polarons have a high mobility; in this case, the energy
of thermal activation of a trap is determined by the energy level,
from which a charge passes to the polaron band; b) if relaxed
polarons are capable of self-trapping, then a thermally excited
charge can tunnel to the neighboring equivalent position in the
lattice on the assumption that there appears the corresponding
configuration of the ionic surrounding caused by thermal motion; c)
if a luminescence center is close to a trap, there appears a
possibility of tunneling of the excited charge directly to the
excited LC state. Thus, variant (a) determines the band mechanism of
transfer of a charge thermally released from an LC, whereas the
other two mechanisms are realized without participation of the band.

It is shown [10] that charges are thermally released from  traps
with the energy multiple $\hbar\omega$, if they tunnel to the
equivalent neighboring crystalline surrounding without passing to
a band state and reach an LC after several such jumps. They can
also directly tunnel to the excited state of a recombination
center if there is one in the neighborhood. If the trap energy is
determined by a half-integer value of the quantum number, then the
charge is released from the trap to the band with the following
recombination at an arbitrarily located LC. Such a mechanism is
realized in all the materials we investigated that do not belong
to AHCs; partially, it takes place in AHCs, where the both
indicated mechanisms are realized. Moreover, they appear equivalent
in NaI, as one can see from Table 2, where integer and
half-integer values \emph{n} alternate.

\subsection*{4.2. Nature of TL-active traps}

Models of the TL phenomenon in AHCs proposed by various authors
reduce to three basic ones: electron, hole, and ion models. Their
analysis carried out in our previous work [10] demonstrated that the
only consistent mechanism is the ion one. In Introduction, we noted
that, in LiF and NaCl, the vibrational quantum energies correlate
with the atomic masses of halides. It turned out that such a
correlation also exists for KCl and NaI (for LiF and KCl:
$\hbar\omega_{\mathrm{LiF}}/\hbar\omega_{\mathrm{KCl}}=1.34$,
$\sqrt{m_{\mathrm{Cl}}/m_{F}}=1.37$; for KCl and NaI -- 1.99 and
1.89; KCl and NaCl -- 1.08 and 1.0; LiF and NaCl -- 1.44 and 1.37;
LiF and NaI -- 2.66 and 2.59; NaCl and NaI -- 1.84 and 1.89,
correspondingly). A certain disagreement between the ratios of the
TL frequencies and the inverse ratios of the roots of the atomic
masses is natural, as the force constants of crystals must somewhat
differ due to the difference in their anion components. Thus, TL
frequencies in AHCs are related to vibrations of halide molecules;
moreover, these molecules are present in each filled trap, as it is
the only way to explain the existence of the oscillatory rule for
trap energies. Such a molecule can be, first of all, presented by an
\emph{H}-center, i.e. an interstitial halide atom that, together
with the nearest site halide ion, forms an $X_{2}^{-}$ molecule
centered at a halide site in [110] direction.\looseness=1

In [20], there were calculated the energies of migration of
interstitial atoms  and vacancies in AHCs. It was established that
the minimal migration energy is characteristic of interstitial anion
atoms -- \emph{H}-centers. For the investigated crystals, these
values lie in the range 0.05--0.16~eV, correlate with the ionization
energies of traps (our data), and determine the minimal trap energy,
whose role is played by the self-trapped state of the
\emph{H}-center.

It is worth noting that a hole in the relaxed state also has a
two-center localization in AHCs and can be considered as a
single-charged halide molecule $X_{2}^{-}$ ($V_{K}$-center).
However, the calculations and experiment [21,22] demonstrate that the
local frequencies related to the $V_{K}$-center in the
investigated crystals are lower than the corresponding TL
frequencies by a factor of 3--4. Thus, we have grounds to state
that TL in AHCs is caused by the release of \emph{H}-centers from
traps with their following recombination.

A thermally released\emph{H}-center can recombine in two ways: at
an \emph{F}-center  and at an anion vacancy. In the first case,
there appears the nodal halide atom at a place of the
\emph{F}-center. In [15--18], it is assumed that the energy of
{\emph{H-F}} recombination is transformed into the energy of radiative
electron transition. However, the radiative efficiency of \emph{H-F}
recombination is extremely low [19]. The model becomes logically
completed if one allows for the recombination of \emph{H}-centers at
anion vacancies [10]. This reaction produces a hole and a nodal
anion, which is followed by the radiative recombination of the hole
($V_{K}$-center) at the \emph{F}-center and the formation of a
halide vacancy. These two successive reactions result in the
generation of a light quantum corresponding to the characteristic
radiation band present in the TL emission spectrum of all traps and
the healing of one \emph{F}-center. In this case, the concentration
of anion vacancies remains constant.

\subsection*{4.3}

 Thus, KCl and NaI have joined the list of
crystalline  materials with the oscillatory rule observed in the
energy spectrum of traps. In KCl, the majority of the energies
appeared to be multiple of an integer number of vibrational quanta
$\hbar\omega_{\mathrm{KCl}}=0.121$ eV. Based on the estimates of the
statistical sum for the oscillatory states, it was concluded that TL
in KCl is mainly realized by means of the charge tunneling from excited
vibrational states of traps to those of luminescence centers without
transition to a band state. That is, the activation energy of traps
appears to be multiple of an integer number of vibrational quanta.
If a charge passes to a recombination center via a band, then the
activation energy corresponds to a half-integer number of
vibrational quanta (there exist three such values in KCl). In NaI,
the value of the vibrational quantum
$\hbar\omega_{\mathrm{NaI}}=0.061$ eV. TL is realized both by means
of the charge tunneling from excited vibrational states of traps to
those of luminescence centers without the transition to a band state and
by means of the charge transition to a recombination center via a band.
The both mechanisms of transition of a charge carrier from a trap to
a luminescence center are equivalent and competitive.

In the earlier investigated materials, the vibrational quantum
energies  coincided with the energies of certain Raman scattering
lines. But, in AHCs, these energies correspond to local
vibrational modes. Based on the existence of the correlation
between the values of $\hbar\omega$ in LiF, NaCl, NaI, and KCl and
the anion masses, these frequencies can be ascribed to a local
vibrational mode of the $X_{2}^{-}$ halide molecule
(\emph{H}-center).

\section{Conclusions}

The existence of the oscillatory
rule in the energy spectrum of traps is  caused by the polaron
effect, and this regularity represents a universal property of
crystals with the mainly ion-bond type. The parameters of this
regularity can be established by means of precise measurements,
whose required accuracy can be provided by the method of
fractional TL due to the possibility of obtaining a considerable
number of single-type data and their deep mathematical treatment.

\noindent -- It is shown that, in KCl and NaI AHCs, the thermal
activation energies of traps form one oscillatory series with the
vibrational quantum energies $\hbar\omega=0.121$ eV (979 cm$^{-1}$)
in KCl and $\hbar\omega=0.061$ eV (492 cm$^{-1}$) in NaI.

\noindent -- The generalization of data concerning the investigated
alkali-halide crystals NaCl, LiF, KCl, and NaI allowed us to confirm
the earlier proposed model of TL in AHCs. In particular, it is
assumed that, in addition to the nonradiative \emph{H-F}
recombination, there exists the two-stage recombination of
\emph{H}-centers at anion vacancies resulting in the radiative
recombination of a hole at an \emph{F}-center.

\noindent -- The found oscillatory rule in the energy spectrum of
traps in KCl and NaI has confirmed the conclusion that this
regularity is caused by the polaron effect and is universal for
crystals with the mainly ion-bond type.

\vskip3mm The authors thank to V.F. Kuznetsov for the considerable
technical assistance in carrying out the experiment.







\end{document}